\documentclass[preprint,review,10pt,sort&compress]{elsarticle}
\usepackage{fullpage,enumitem,amsmath,amssymb,graphicx,xcolor}
\usepackage[margin=1.0 in]{geometry}
\usepackage{multirow}
\usepackage{cases}
\usepackage{float}
\usepackage{xcolor}
\usepackage{url}
\usepackage{comment}
\usepackage{natbib}
\usepackage{amsmath}
\usepackage{nomencl}
\makenomenclature

\begin{document}
\begin{frontmatter}

\author[address]{Mircea Cozmei}
\author[address]{Tristan Hasseler}
\author[address]{Everett Kinyon}
\author[address]{Ryan Wallace}
\author[address]{Antonio Alessandro Deleo}
\author[address]{Marco Salviato\corref{cor2}}

\address[address]{William E. Boeing Department of Aeronautics and Astronautics, University of Washington, Seattle, Washington 98195, USA}

\title{Aerogami: Composite Origami Structures as Active Aerodynamic Control}

\cortext[cor2]{Corresponding Author, \ead{salviato@aa.washington.edu}}

\begin{abstract}
This study explores the use of origami composite structures as active aerodynamic control surfaces. Towards this goal, two origami concepts were designed leveraging a combination of analytical and finite element modeling, and computational fluid dynamics simulations. Wind tunnel tests were performed at different dynamic pressures in conjunction with two different active control laws to test the capability of obtaining desired drag values. The experiments revealed excellent structural rigidity and folding characteristics under aerodynamic loading. These results are in very good agreement with one-way Fluid Structure Interaction (FSI) simulations, which show the potential of using high-fidelity modeling for the design and optimization of these structures.

Future work will focus on developing advanced origami designs that allow for more deterministic folding as well as improved weight, stiffness, and fatigue characteristics in the use of materials.
\end{abstract}

\end{frontmatter}

\mbox{}
\nomenclature{$x(\alpha)$}{Displacement of the actuator}
\nomenclature{$\alpha$}{Angle of attack of the front plate}
\nomenclature{$l_{FP}$}{Length of the front plate}
\nomenclature{$l_{BP}$}{Length of the back plate}
\nomenclature{$\psi$}{Dimensionless ratio of the lengths  $l_{BP}$/$l_{FP}$}
\nomenclature{$D$}{Drag Force}
\nomenclature{$D(\alpha)$}{Drag Force as a function of angle of attack}
\nomenclature{$C_D(\alpha)$}{Coefficient of Drag as a function of angle of attack}
\nomenclature{$q$}{Kinetic energy per unit volume of the fluid}
\nomenclature{$\rho$}{Fluid density}
\nomenclature{$v$}{Fluid velocity}
\nomenclature{$S$}{Surface area of the origami}
\nomenclature{$\omega$}{Depth of the whole origami structure}
\nomenclature{$\textbf{w}$}{Geometrical vector of the edge of the Dino origami fan}
\nomenclature{$\theta$}{Angle between the edge vectors of the Dino origami fan}
\nomenclature{$E$}{Young's modulus}
\nomenclature{$\nu$}{Poisson's ratio}

\printnomenclature[3cm]

\section{Introduction}

Composite materials continue to be an important and active area of research in materials science and engineering. This continued research is due in large part to the significant benefits over traditional metal/alloy structures, including weight savings, superior tailorability, and superior specific strength and fatigue resistance compared to other materials \cite{baker,Ahmed,kaiser,Kar03, Brandt08, Ceccato17, Carloni16, Feo13,sorensen,chortis, Yao1, Yao2, Quaresimin1}. However, despite the several benefits, the problem of readily and efficiently constructing more complex shapes with composite materials remains \cite{lin2014composite}. Recent works at the University of Washington have explored the feasibility and possible advantages of combining the ancient art of origami and composite materials \cite{deleo2018composite, o2018deployable}. The resulting systems feature configurations of many smaller pieces of composite laminates that are easily constructed in bulk which fit together in intricate shapes using a variety of flexible joints, allowing streamlined construction of highly complex geometries with composite materials. 

In recent years, the interest in thick foldable composites has been continuously increasing \cite{chil20}. Boncheva and Whitesides \cite{boncheva2005templated} investigated the manufacturing of foldable structures via the coating of paper origami by an elastomer. Thanks to the compliant design, the structures could be deployed leveraging a relatively simple pneumatic system. However, the resulting structural strength was inevitably low. On similar grounds, an elastomeric, foldable system reinforced by carbon fibers was proposed and analyzed in \cite{chillara2018stress} to enable the achievement of smooth folds in lightweight structures. In this case, the actuation could be provided by active laminae such as e.g. fluidic layers or through passive means such as a pair of external forces.

Taking inspiration from Nature, Daynes \emph{et al}. \cite{daynes2014bio} demonstrated a system consisting of multiple cells made from silicone rubber with locally reinforced regions based upon Kirigami principles. Thanks to the existence of a structural bistability, each cell was shown to be able to maintain its shape in either a retracted or deployed state without the aid of mechanisms or sustained action. A very interesting foldable system called "fluidic origami" was proposed in \cite{li2015fluidic} by Li and Wang who explored the use of thin steel facets connected by an adhesive to enable the design of multi-stable structures inspired by the plant kingdom. Thanks to the combination between strong and stiff materials and multi-stability, the concept displayed outstanding tunability and functionalities. Several other interesting self-foldable designs where the actuation leverages e.g. electric or magnetic fields or shape memory alloys can be found in the literature \cite{lan2009fiber, von2013folding, ahmed2013multi, peraza2013design, felton2014method, ahmed2017electric}. 

The foregoing systems can pave the way to model and control the actuation and kinematics of complicated structures, ranging from morphing wings \cite{seigler2007modeling, barbarino2011review}, to harvesting power from vibration using piezoelectric materials \cite{sodano2004review}. An outstanding review of the most recent developments in foldable composites for applications in morphing structures can be found in \cite{chil20}.

Motivated by the great potential for the attainment of morphing systems, the purpose of this project is to explore the use of origami composites as aerodynamic control surfaces. Different from other designs, the concept investigated in this study, called "Aerogami", makes use of lightweight, strong, and stiff thermoset matrix carbon fiber composites both for the creases and the rigid facets. The purpose is to develop novel designs with unprecedented mechanical and aerodynamic performance combining the outstanding specific mechanical properties of composites and the tunability of origami. While movable structures have been used as aerodynamic control surfaces \cite{wilde1998aerodynamic, grosjean1998micro, sanders2003aerodynamic} in the past, the inherent advantages of lightweight and strong composite structures have not been explored yet. The origami technology proposed in this study could enable the development of large drag surfaces that can be stowed in a much smaller area than traditional airbrakes which are currently used particularly in the competitive automotive sector such as Formula 1 \cite{lofdahl2018evolution}.

Additionally, this project seeks to address current challenges with actuation through design of origami shapes that can be controlled via a single linear actuator. Active feedback control systems \cite{ogata2002modern} were implemented into the design, allowing the aerodynamic surface to automatically adjust shape given varying airflow conditions in order track a desired drag force.

Wind tunnel testing was performed on two original origami designs in the University of Washington $3$ ft $\times$ $3$ ft wind tunnel with the following goals: (i) to measure drag forces at different actuation percentages and dynamic pressures, (ii) to test control laws and their ability to track a given drag value, and (iii) to observe structural integrity and flutter characteristics of the test assembly. These tests were run at varying actuation percentages (10-90\%) at dynamic pressures ranging from 479 to 1915 $\mbox{ Pa}$ (corresponding to 10-40 psf). Finally, a one-way Fluid Structure Interaction (FSI) was performed to check and confirm the drag values calibrated by the load cell and the overall fluid-structure behavior.


\section{Theoretical framework}

\begin{figure}[H]
        \centering
        \includegraphics[width=1\linewidth,angle=0]{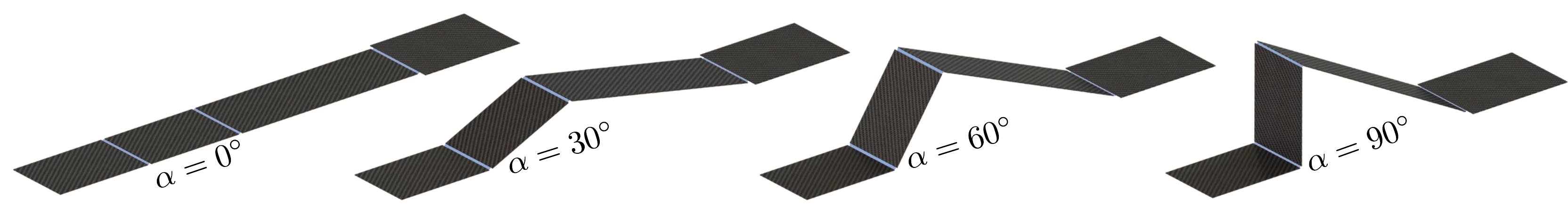}
        \caption{Worm retracted/extended}
        \label{fig:CONFIG1}
    \end{figure}

\begin{figure}[H]
        \centering
        \includegraphics[width=1\linewidth,angle=0]{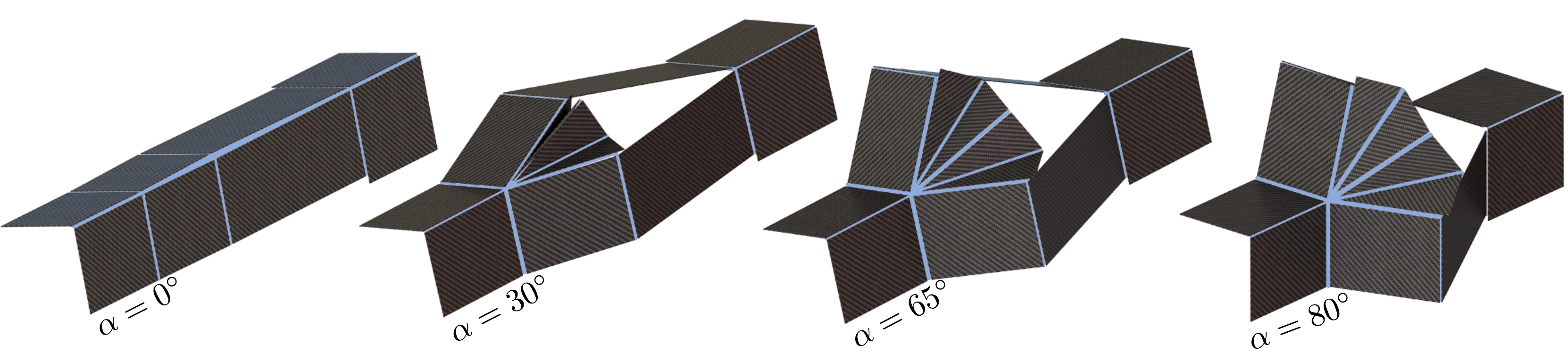}
        \caption{Dino deployment stages}
        \label{fig:CONFIG2}
    \end{figure}

The goal of the present study is to demonstrate the potential use of composite origami as drag control surfaces. Considering the exploratory spirit of the investigation, two different geometries of origami structures were designed: Worm, a simple four-panel design as shown in Fig.~\ref{fig:CONFIG1}, and Dino, a more complex configuration incorporating two Worms connected by a central fan inspired by the frilled lizard (Fig.~\ref{fig:CONFIG2}). A notable advantage of the foregoing origami surfaces compared to traditional control surfaces is that, notwithstanding the complex folding behavior, they require only a linear actuation system.

The choice of analyzing these configurations was made to favor simplicity and robustness of the folding mechanism and of the actuator while providing a good aerodynamic performance. The "Worm", represents the simplest configuration featuring all the main folding characteristics of a structural origami while providing good drag control. As such, this topology was considered the perfect candidate to demonstrate the main features of Aerogami. "Dino" was chosen to explore the effects of a slightly more complex topology on both the aerodynamic performance and deployability.

It is worth mentioning here that the technology presented in this work enables origami composites way more complicated than "Worm" and "Dino". The additional complexity of the design may be justified by the particular boundary conditions, and performance requirements. In such a case, however, the best topology should be identified via a thorough optimization study. This is beyond the scope of the present work and will be the subject of future publications.

The following sections briefly describe the folding kinematics and drag forces of Worm and Dino.

\subsection{Folding Kinematics}
\begin{figure}[H]
    \centering
    \includegraphics[width=1\linewidth,angle=0]{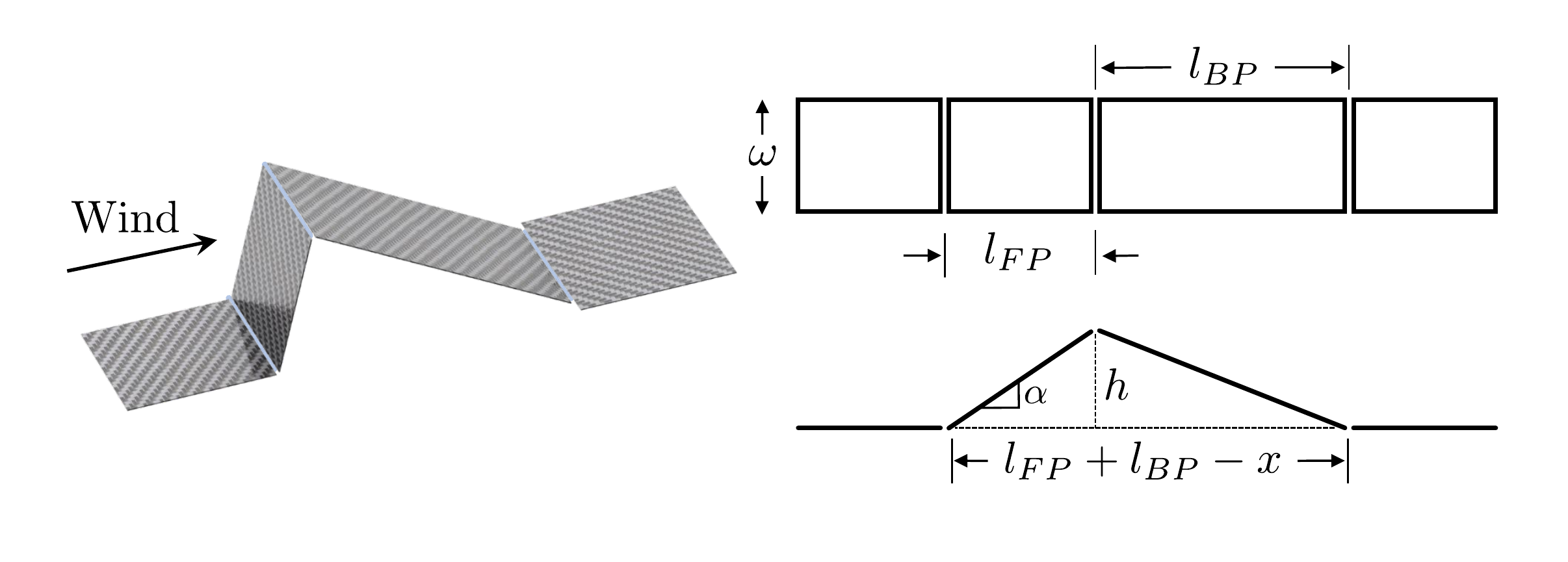}
    \caption{Geometry used in derivation of $X(\alpha)$}
    \label{fig:geometry}
\end{figure}

\begin{figure}[H]
    \centering
    \includegraphics[width=0.7\linewidth,angle=0]{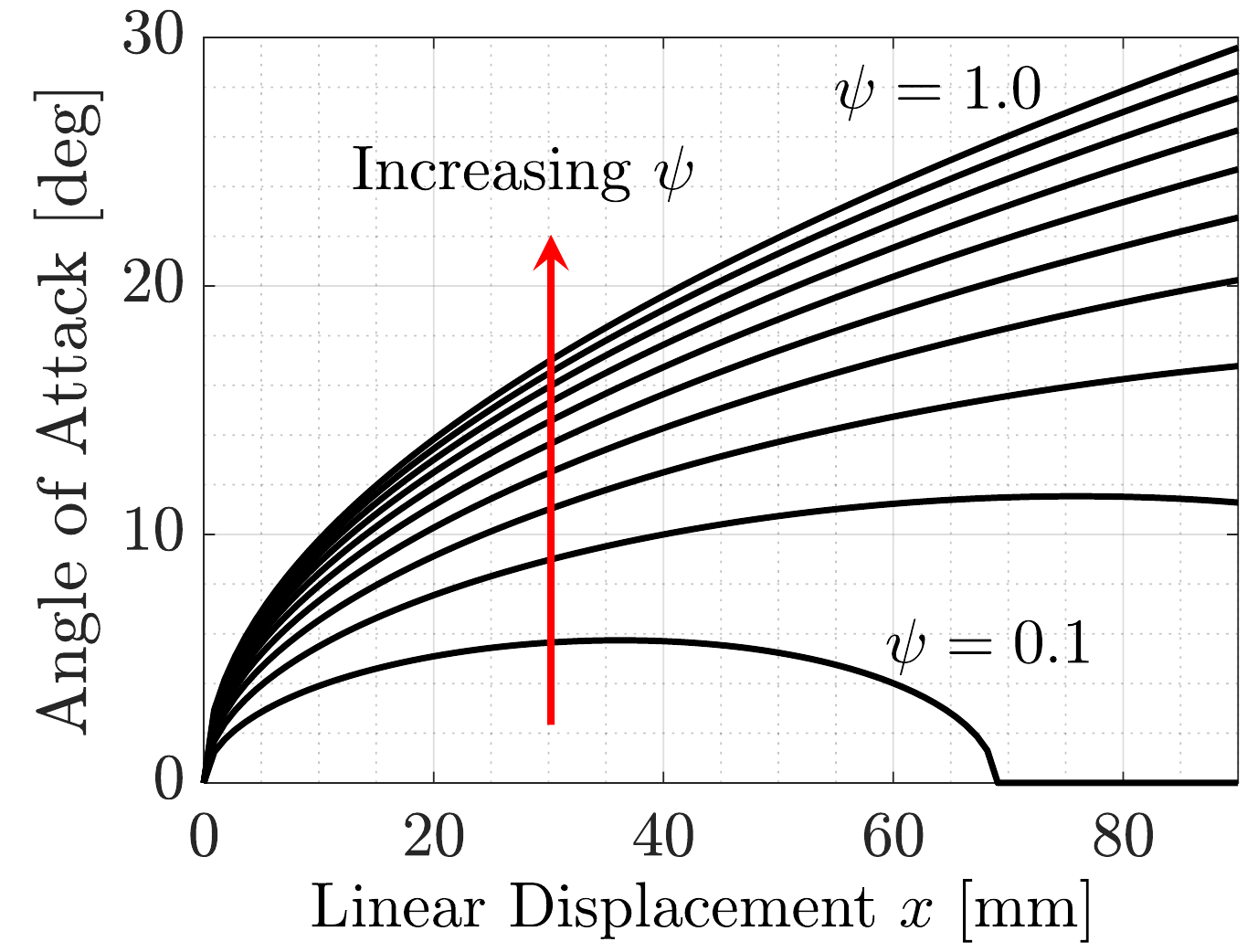}
    \caption{Plot of the angle of attack as a function of the linear displacement for increasing values of $\psi$}
    \label{fig:psi_plot}
\end{figure}

The displacement of the actuator $x(\alpha)$ as a function of the angle of attack of the front plate, $\alpha$, of Worm and Dino can be derived from the geometry in \textbf{Fig.~\ref{fig:geometry}}. The inverse of this function $x^{-1}(\alpha) = \alpha(x)$ can then be computed analytically, giving the following simple equation for the angle of attack as a function of actuation displacement: 

\begin{equation}
\alpha(x) = \arccos \left[ \frac{l_{BP}^2 - l_{FP}^2 - (l_{FP} + l_{BP} - x)^2 }{2l_{FP}(x - l_{FP}- l_{BP})}  \right] = \arccos \left[ \frac{\psi^2 - 1 - (1 + \psi - x/l_{FP})^2 }{2(x/l_{FP} - 1 - \psi)}  \right]
\label{Xalpha}
\end{equation}

where the dimensionless geometrical parameter $\psi=l_{BP}/l_{FP}$ was introduced. As can be noted from \textbf{Fig.~\ref{fig:psi_plot}}, which shows the angle of attack as a function of the linear displacement $x$, the foregoing relationship is highly nonlinear. Proper design of the ratio between the lengths of the frontal and back panel can be used to obtain the desired evolution of the angle of attack during folding.

\subsection{Worm Drag Analysis}

The drag force, $D$ generated by the origami surface can be generally expressed as: 

\begin{equation}
D\left(\alpha\right) = C_{D}\left(\alpha\right)qS
\label{drag}
\end{equation}

where  $q=1/2 \rho v^2$ represents the kinetic energy of the fluid, $\rho$ is the fluid density, $v$ is the flow velocity, $S$ is the surface area of the origami, and $C_D$ is a drag coefficient which can be found in aerodynamics handbooks \cite{anderson2010fundamentals} or calculated by Computational Fluid Dynamics (CFD)\footnote{A note for the reader, the superscripts $\mbox{w}$ and $\mbox{d}$ will refer to the Worm and Dino respectively.}. By simple trigonometric arguments, the frontal projection of the surface area $S$ of Worm can be written directly as a function of the linear displacement:

\begin{equation}
S^{\text{w}} (x) = l_{FP} \cdot \sin{\left [\alpha(x)\right]}\cdot \omega
\label{surface}
\end{equation}

where, as shown in \textbf{Fig.~\ref{fig:geometry}}, $\omega$ is the plate width and $\alpha(x)$ is used from \textbf{Eq.~\ref{Xalpha}}.
Finally, combining \textbf{Eqs.~\ref{Xalpha}}, \textbf{\ref{drag}}, and \textbf{\ref{surface}}, it is possible to express the drag force in Worm as a function of the displacement of the linear actuator, $x$, by the following simple expression:

\begin{equation}
D^{\text{w}}\left(x\right) = C_D\left[\alpha(x)\right]q \cdot S^{\text{w}}(x)
\label{Dda_worm}
\end{equation}

\subsection{Dino Drag Analysis}
\begin{figure}[H]
    \centering
    \includegraphics[width=1.0\linewidth,angle=0]{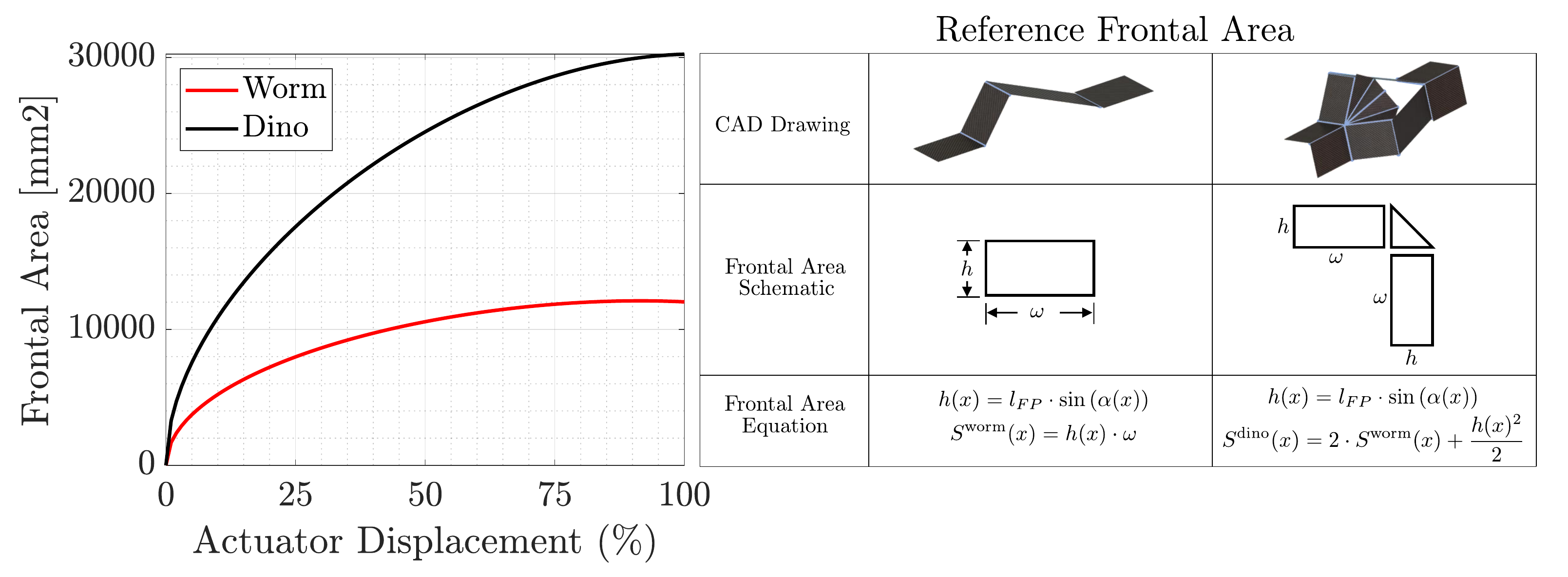}
    \caption{Area difference between the Worm and Dino as a function of the Actuator Displacement.}
    \label{fig:area_difference} 
\end{figure}

The folding kinematics for the Dino composite origami is the same as that of two orthogonal Worms with the addition of a fan in-between. Thus, the main addition to the drag $D^{\text{d}}(x)$ is related to the unfolding of the fan which can be simply approximated as a triangular area between the frontal projections of the two Worm plates, as graphically shown in \textbf{Fig.~\ref{fig:area_difference}}.

\begin{figure}[H]
    \centering
    \includegraphics[width=1\linewidth,angle=0]{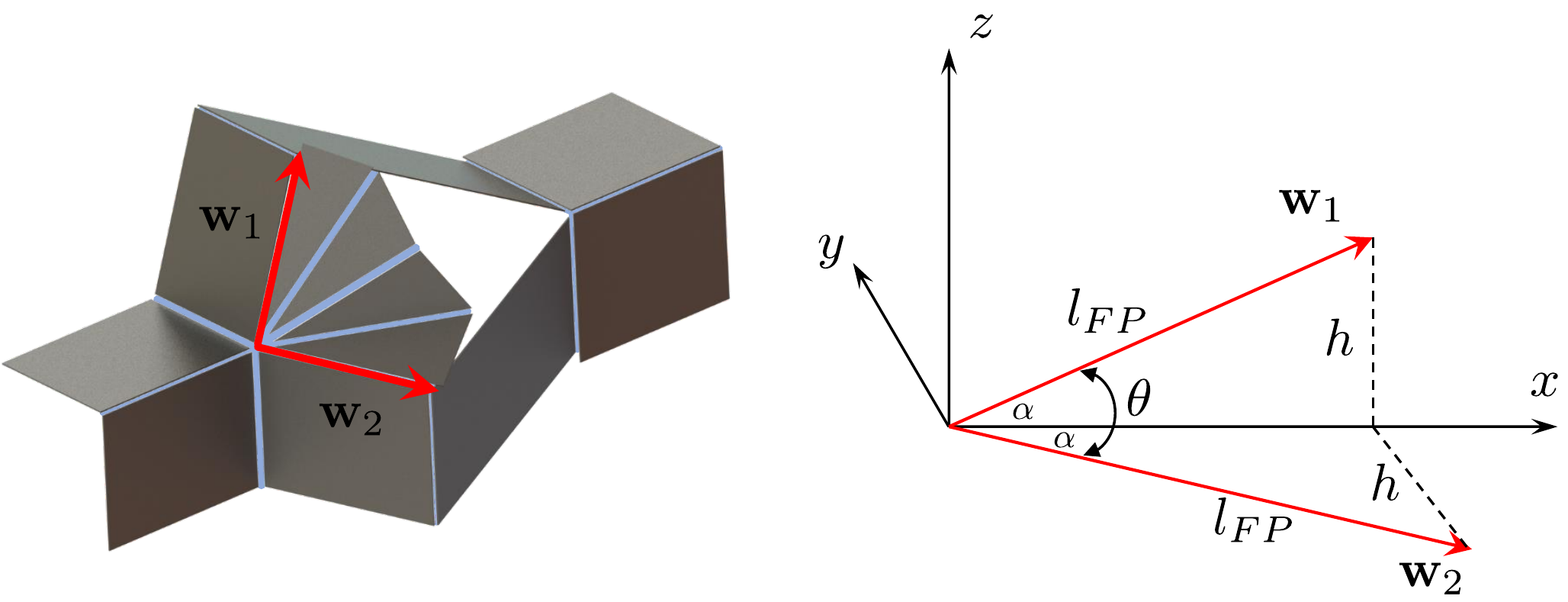}
    \caption{The edges of the fan on Dino represented geometrically as vectors $\textbf{w}_1$ and $\textbf{w}_2$.}
    \label{A6}
\end{figure}

With reference to the schematic in \textbf{Fig.~\ref{A6}}, where the edges of the fan are given by two vectors $\textbf{w}_1$ and $\textbf{w}_2$, one can write the following equations:

\begin{equation}
\textbf{w}_1 = \left[ l_{FP}\cos{(\alpha)} , 0 , l_{FP}\sin{(\alpha)} \right]^T 
\end{equation}

\begin{equation}
\textbf{w}_2 = \left[l_{FP}\cos{(\alpha)}, -l_{FP}\sin{(\alpha)}, 0 \right]^T 
\end{equation}

The angle $\theta$ between these vectors is the angle at which the fan is deployed. Noting that $\textbf{w}_1 \cdot \textbf{w}_2 = (l_{FP})^2 \cos(\theta)$, this angle can be determined from the scalar product between the $\textbf{w}_1$ and $\textbf{w}_2$. This gives the following equation:

\begin{equation}
\theta = \cos^{-1}(\cos^2(\alpha))   
\end{equation}

Leveraging the foregoing expression, the surface area as a function of the linear displacement, $x$, can be calculated as follows:

\begin{equation}
 S^{\text{d}}(x) = 2l_{FP} \sin{(\alpha(x))} \cdot \omega + \frac{l_{FP}^2\sin^2{(\alpha(x)})}{2}
\label{DinoSurfaceArea}
\end{equation}

where $w$ is the width of the front plate (Fig. \ref{fig:area_difference} compare the surface areas of Worm and Dino and show references schematic for the frontal areas and $h = l_{FP} \cdot \sin{\alpha(x)}$).

Then, combining \textbf{Eqs.~\ref{Xalpha}}, \textbf{\ref{drag}}, and  \textbf{\ref{DinoSurfaceArea}}, it is possible to express the drag force in Dino as a function of the displacement of the linear actuator, $x$, by the following simple expression:

\begin{equation}
D^{\text{d}}\left(x\right) = C_D\left[\alpha(x)\right]q \cdot S^{\text{d}}(x)
\label{Dda_dino}
\end{equation}

\section{Design of the origami control surfaces}
The design of the origami surfaces for their deployment was based on the coupling between Computational Fluid Dynamics (CFD) and Finite Element (FE) simulations into a one-way Fluid Structure Interaction (FSI) framework. The initial structural design was tested using ANSYS Fluent \cite{Fluent} and the aerodynamics results used as input in the structural solver of ANSYS Mechanical and subsequently re-run with Ansys Fluent with the new deformed geometry to allow convergence of the quantities of interest (drag and displacement). 

The initial structure was designed by assuming the greatest loading scenario: the front plate fully perpendicular to the free stream flow and with a flow velocity of $53.6$ m/s ($120$ mph). The materials used were the Toray T700 12k Plain Weave CFRP with the following mechanical properties: $E_1 = 85 \mbox{ GPa}$ and $E_2 = 85 \mbox{ GPa}$, a shear modulus of $G_{12} = 5 \mbox{ GPa}$, and a Poisson's ratio of $\nu_{12}$ = $0.1$ \cite{Toray}; and GFRP impregnated with urethane epoxy \cite{Sharkthane} with the following mechanical properties: $E_1 = E_2 = 1.93 \mbox{ GPa}$, a shear modulus of $G_{12} = 3.40 \mbox{ MPa}$, and a Poisson's ratio of $\nu_{12}$ = $0.21$ \cite{deleo2018composite}. A preliminary simulation obtained using FEMAP \cite{Femap} showed that a 6-ply laminate sandwiched with a mid fiber glass layer previously cured using the aforementioned urethane epoxy, with all the plies aligned in the longitudinal direction of the structure yielded a maximum displacement of less than $1.0$ mm and very low failure indexes, which was considered acceptable and therefore utilized for all future laminates. The control laws for the aerogami deployment are discussed at the end of this section after the aero-structural part.

\subsection{One-Way Fluid Structure Interaction (FSI) simulations}
\begin{figure}[H]
        \centering
        \includegraphics[width=1\linewidth,angle=0]{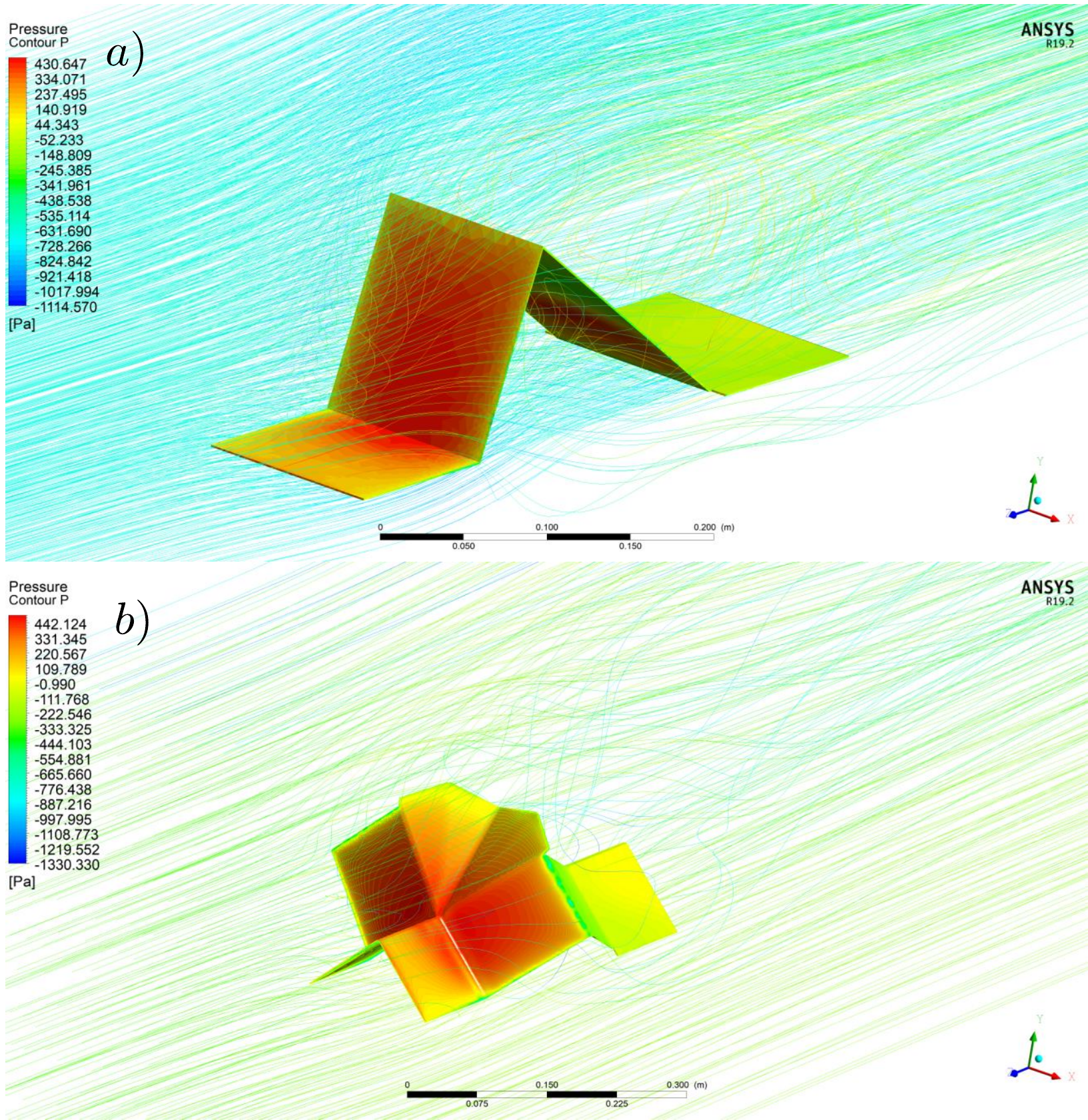}
        \caption{Fluent Results of a) Worm Aerogami, b) Dino Aerogami.}
        \label{fig:FSI_results}
    \end{figure}

The proposed origami surfaces must provide desired values of drag force while guaranteeing structural integrity. To obtain a proper understanding of the aerodynamic forces and to optimize the design of the proposed concepts, CFD analyses coupled with FEA were conducted. \textbf{Fig.~\ref{fig:FSI_results}} shows the velocity streamlines of the Worm and Dino at the 90\% and 70\% actuation configurations respectively for a free-stream velocity of $26.8 \mbox{ m/s}$ ($60 \mbox{ mph}$) calculated by ANSYS Fluent \cite{Fluent}, using the k-$\omega$ SST turbulence model \cite{FluentSST}. This model was chosen for its accuracy in simulating separated flow, an aspect that is critically important for the design of the origami control surfaces. In the simulation, the density of air was assumed to be $1.225$ kg/m$^{3}$ (sea level density) and the temperature was assumed to be $20^{\circ}$C. The Reynolds number was estimated to be $900,000$ assuming a chord length of $0.5 \mbox{ m}$ and a dynamic viscosity of $1.821 \cdot 10^{-5} \mbox{m}^2$/s (at 20$^{\circ}$C). It is important to note that the full testing assembly was not run in CFD in order to isolate the drag produced solely by the origami. The drag outputted by CFD was $9.64 \mbox{ N}$ for Worm and $26.8 \mbox{ N}$ for Dino. The projected areas were $0.0148$ m$^2$ for Worm and $0.0434$ m$^2$ for Dino. Therefore, the Dino produced 278\% more drag with a 293\% larger deployed surface area. For $\alpha=90^{\circ}$, $C_D$ $\approx$ $1.2$ \cite{anderson2010fundamentals} and \textbf{Eq.~\ref{Dda_worm}} gives a drag force of $45 \mbox{ N}$. All the drag values were imported into the mechanical solver using the coupling system toolbox of Ansys Workbench which took care of exchanging data between the two different solvers.

Regarding the structural Finite Element Analyses, the geometry used was exactly the same and a set of interaction surfaces was defined in order to have information of displacement and boundary forces interchanging between the structural solver (ANSYS Mechanical) and the fluid solver (ANSYS Fluent). For the Worm, $155200$ solid brick linear elements were used with orthotropic mechanical behavior calculated using Classical Lamination Theory (CLT) with the base properties listed at the beginning of this section. For the Dino $1743145$ elements were used with the same properties. The urethane flexible joints between the rigid CFRP facets were modeled as partitions of the solid bodies at the edges between the facets.

As the following sections will discuss, although a Coupled Eulerian Lagragian framework could have provided a better approximation of the interaction between the fluid and the Aerogami, the results of the simulations with one-way FSI were in very good agreement with the experimental data obtained in the wind tunneland were able to capture even the effects on the drag force of the deflections of the fan in Dino.

\subsection{Controller Selection}
Using the results found from analytical derivations, computations, and the testing of control hardware, a Simulink model was developed to assist in control design. With this model in place, two control laws were designed taking into account the noise levels expected in the load sensor (less than $\pm$0.1 N) and the modeled behavior of the Arduino and the linear servo. Based on simulation results, a proportional \cite{ogata2002modern} control law was chosen for software implementation. A "bang-bang" \cite{bellman1956bang, ogata2002modern} control law was also explored but was not robust enough to be implemented. This is because a bang-bang controller is a feedback controller that can only output two states. In this case: maximum deployment rate and maximum stow rate. These two outputs were governed by how the measured force was related to the desired force (greater than or less than). A proportional controller, on the other hand, is a feedback control law that sets the actuation response directly proportional to the error between measured state and desired state. The constant of proportionality is called a “gain” which was tuned to provide maximum system efficiency.

The motivation for either of these two presented choices is threefold. First, the servo system adopted in this work was selected to favor robustness, simplicity, and cost-effectiveness over speed. Accordingly, a controller with low rise time was desired to mitigate the slow actuator movement. Second, the servo could only receive commands that set an extension length, making it difficult to control the velocity of the actuation. Third, the existence of only one sensor meant that PID control laws would require a large amount of sensor processing to numerically approximate the velocity and integral of the state potentially adding additional delay and inaccuracies.

Lastly, it is worth mentioning that the inertial terms that often lead to underdamped behavior in proportional controllers \cite{ogata2002modern} were found to be negligible in the Worm and Dino due to the low weight of the composite structure. Through simulations, the proportional controllers displayed adequate behavior for the available hardware.

\section{Manufacturing}
The fabrication of the composite origami consisted of several steps, which followed the manufacturing procedure recently proposed by the authors in \cite{deleo2018composite} and \cite{o2018deployable} using Vacuum Bag Only (VBO). First, pre-impregnated Toray T700 12k Plain Weave CFRP plies \cite{Toray} were taken out of -11$^{\circ}$C storage, let thaw in a vacuum bag, and placed in a CNC fabric cutter by Autometrix \cite{Autometrix}. Then, the desired panels were precisely cut out and curing of the thermoset resin followed. 
Two metal plates were cleaned using scrapers and acetone before a thin layers of release agent was applied to the plates. Three plies 
of carbon panels were combined into a $[0^\circ/0^\circ/0^\circ]$ layup for increased strength. These panels were placed between the two metal plates and cured in the hot press following the manufacturer specifications. 
Teflon sheets were used in between the carbon panels and the metal plates as needed to ensure easy separation after curing. The composite panels were then cut with a water diamond saw and sanded down to their precise dimensions.

\begin{figure}[H]
        \centering
        \includegraphics[width=1\linewidth,angle=0]{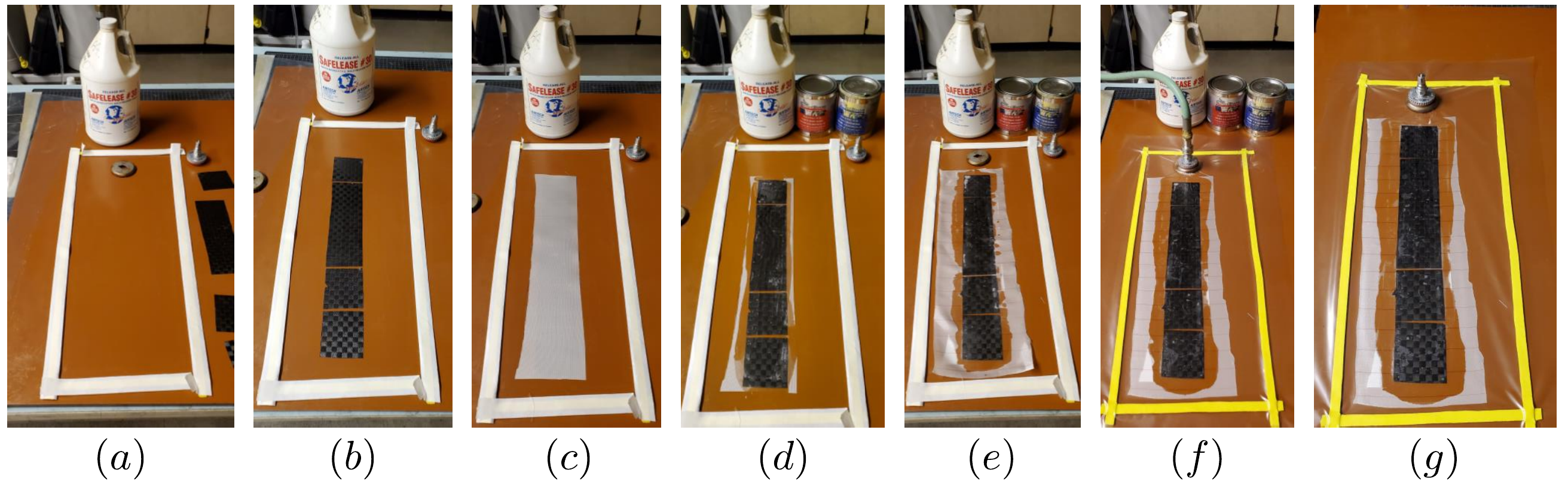}
        \caption{Step by step manufacturing of the Worm panel using VBO approach.
        (a) A working surface is cleaned and release agent is applied.
        (b) The first layer of the facets is laid down.
        (c) The dry fiberglass sheet is deposited.
        (d) The urethane epoxy is manually spread over the fiberglass.
        (e) The top layer of facets is laid down.
        (f) The teflon or peelply is laid on top of the specimen and the vacuum bag is pushed down.
        (g) The specimen is let curing overnight.}
        \label{fig:VBO_steps}
    \end{figure}

 
Once the composite panels were prepared, the next stage was to prepare a sandwich structure were two sides of CFRP facets are joined together using a single layer of dry fiberglass in between impregnated with a very flexible urethane epoxy provided by Sharkthane \cite{Sharkthane} with A shore number of $30$. To do so, a small amount of glue was temporarily used to correctly lay down the Dino and Worm configurations on a large metal or garolite plate, as shown in \textbf{Fig.~\ref{fig:VBO_steps}}a-b. Bagging Tape was then placed around the panels and a sheet of dry fiberglass was placed on top of the panels (\textbf{Fig.~\ref{fig:VBO_steps}}c). The urethane system was mixed and evenly distributed before placing a set of panels over the fiberglass (\textbf{Fig.~\ref{fig:VBO_steps}}d). A Teflon sheet was placed over the panels and the bridge piece was placed near the edge of the Bagging Tape (\textbf{Fig.~\ref{fig:VBO_steps}}e). The vacuum bag was placed over the Teflon and pushed down into the Bagging Tape to create an airtight seal (\textbf{Fig.~\ref{fig:VBO_steps}}f). The vacuum pump hose was attached to the bridge piece (\textbf{Fig.~\ref{fig:VBO_steps}}g), and the vacuum was turned on and left on overnight. The panels were then taken out of the vacuum bag assembly and trimmed to the right dimensions.

\section{Testing}
\subsection{Experimental setup}

The composite surfaces described in the previous sections were tested to verify their performance in terms of aerodynamics, controllability, and mechanical behavior. All the tests were performed in the $3 \times 3  \times 8$ ft wind tunnel at the University of Washington which is a open-loop facility capable of 135 mph (60 m/s) flows with a $9:1$ contraction ratio.



\begin{figure}[H]
        \centering
        \includegraphics[width=1\linewidth,angle=0]{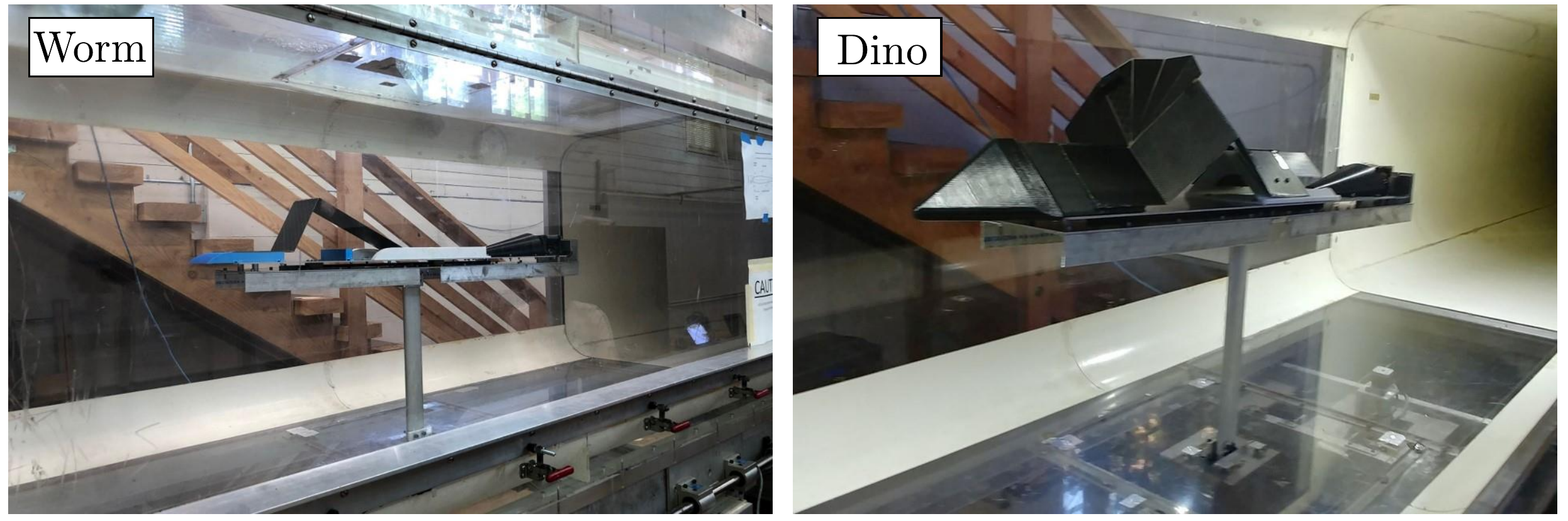}
        \caption{The Worm (left) and the Dino (right) assembly in the 3' x 3' wind tunnel.}
        \label{fig:SW_Rendering}
    \end{figure}

\begin{figure}[H]
        \centering
        \includegraphics[width=1\linewidth,angle=0]{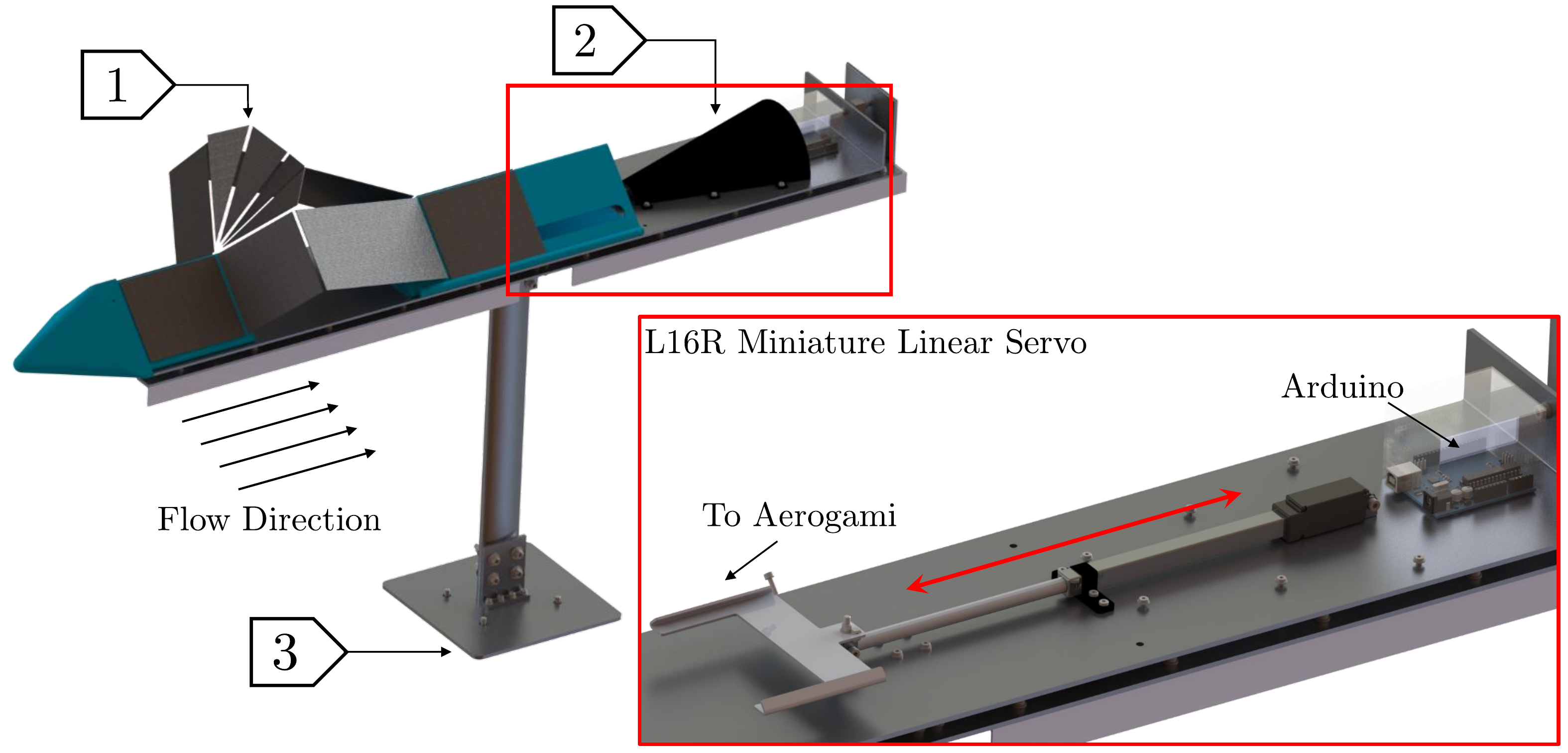}
        \caption{CAD drawing and rendering of the Dino assembly as reference. 1) Dino origami sits on a rail allowing for motion along flow direction, 2) wind shield cover protects linear actuator, Arduino, and load sensor, 3) assembly bolted to wind tunnel floor. On the bottom right a detailed view of the L16R Miniature Linear Servo system.}
        \label{fig:Wind_Tunnel}
    \end{figure}

For the testing, a new origami test stand shown in \textbf{Fig.~\ref{fig:SW_Rendering}} and \textbf{Fig.~\ref{fig:Wind_Tunnel}} was designed to minimize possible airflow perturbations seen by the origami control surface during actuation. In the fully-closed (flat) position, the origami was designed to sit $1.5$ ft above the wind tunnel floor, exactly in the center of the wind tunnel. This minimized boundary layer effects from the wind tunnel walls. As can be noted from \textbf{Fig.~\ref{fig:SW_Rendering}} and \textbf{Fig.~\ref{fig:Wind_Tunnel}}, the test stand was fixed to the floor through four bolts, and the bottom plate supporting the entire test assembly was connected to floor by a center strut with a low drag airfoil cross-section.

During the tests, the origami structure sat on 3D printed PLA components which were bolted to an aluminum plate. All forward facing 3D printed components were designed with contoured leading edges to improve airflow quality over the origami test article. The front origami panel was bonded in place with methyl-methacrylate adhesive to the front fairing.

At the rear of the aluminum plate, an Actuonix L16-R linear actuator \cite{Actuonix} pushed the rear plate of the origami thus actuating the entire control surface. The rear plate of the origami was wider than the other origami plates and slid along two 3D printed rails which were bolted to the aluminum plate. Behind the actuator sat an enclosed Arduino control unit \cite{Arduino} used to give the actuator commands. To protect the control electronics from the flow, a 3D printed conical wind shield was placed over the actuator and in front of the Arduino control box (\textbf{Fig. \ref{fig:SW_Rendering}} arrow $2$). 

The aluminum plate was bolted to a linear rail upon which it slides. The linear rail was then bolted to a lower aluminum plate bolted to the supporting strut. At the rear of the test article a force sensor was placed and bolted to both plates. As the upper plate rested on a rail, its motion was unconstrained along the flow direction, thus loading the force sensor in shear, which consequently measured the drag generated by the origami and the supporting top aluminum plate.



Wind tunnel q-calibration was conducted before any data was taken by sweeping the dynamic pressure from $0$ to $1915 \mbox{ Pa}$ (corresponding to a wind speed of $180$ mph) so that the dynamic pressure value read by the wind tunnel Pitot-static probe matched the true dynamic pressure value. Next the assembly's structural integrity was tested by slowly increasing the dynamic pressure from $0$ up to $1915 \mbox{ Pa}$.

An important aspect of the experimental campaign was the development of the actuation and data acquisition systems. These were designed so that the apparatus would be able to independently read and log forces generated by the origami, to independently execute the controls algorithm, and to fold/unfold the origami shapes via a single linear actuator. An Adafruit data-logging shield \cite{Adafruit} was used to implement data logging capabilities in an Arduino Uno, which served as the main controller. A $20 \mbox{ kg}$ Phigits shear load cell (CZL635) \cite{Phidgets} was used to measure the drag generated by the origami. The $20 \mbox{ kg}$ model was chosen to place expected drag readings in the middle of the sensor's range, improving accuracy. A Spark-Fun load cell amplifier (HX711) \cite{Sparkfun} was used to convert analog to digital readings for the load cell. An Actuonix RC linear servo (L-16R) \cite{Actuonix} was used as the control actuator. This model was chosen for its built-in RC controller, allowing it to be directly controlled by the Arduino unit. 


\subsection{Worm Wind Tunnel Testing}
To characterize the folding behavior and aerodynamics of Worm, a series of actuation sweeps were conducted at $478.8$, $1436.44$, and $1771.6 \mbox{ Pa}$. The sweeps actuated the origami from $10$ to $90\%$ actuation in $10\%$ increments, recording the forces generated by the Worm at each increment. Next the proportional control law was tested multiple times at dynamic pressures of $718.2$, $1436.4$, and $1915.2 \mbox{ Pa}$ targeting drags of $9.81$, $19.6$, and $35.3 \mbox{ N}$ (corresponding to load cell readings of $1$, $2$, and $3.6 \mbox{ kg}$) respectively.


\subsection{Dino Wind Tunnel Testing}

\noindent The wind tunnel testing process for the Dino proceeded in the same way as that of the Worm, albeit with different operating dynamic pressures and drag targets. Because the Dino has a larger surface area than the Worm, the tested dynamic pressures were reduced in an attempt to mitigate the larger stresses developed in the testing apparatus. Therefore, the actuation sweeps were performed from 10 to 70\% actuation in 10\% increments at dynamic pressures of $440.5$, $785.2$, and $991.1 \mbox{ Pa}$ (corresponding to wind speeds of 60, 80, and 90 mph respectively). To test the proportional control responses, controlled deployments were performed at dynamic pressures of $440.7$, $783.4$, and $991.5 \mbox{ Pa}$ targeting drag forces of $14.7$, $24.5$, and $29.4 \mbox{ N}$ (corresponding to load cell readings of $1.5$, $2.5$, and $3 \mbox{ kg}$ respectively). The design of the Dino testing configuration was the same as the Worm, with the exception of the 3D printed parts conforming to its specific geometry. The full Dino testing apparatus can be seen installed inside the 3ft x 3ft wind tunnel in \textbf{Fig.~\ref{fig:Wind_Tunnel}}.


\section{Results and Discussion}
\subsection{Preliminary mechanical investigation}
Before performing the experimental campaign, preliminary tests on Worm prototypes were conducted to verify the structural integrity of the system and identify possible design criticalities. After the first Worm prototype was manufactured, it was concluded that the joint spacing was too large, leading to structural instability. This raised concerns of destructive flutter occurring during the wind tunnel test. In response, a Worm and Dino with smaller joints of $3 \mbox{ mm}$ were manufactured and flutter was not observed during wind tunnel testing. However, wind tunnel results exhibited that the spacing could have been even smaller. At low actuation percentages for the Worm, actuator extension caused the soft joints to buckle and additional actuator displacement was required before the origami would begin folding as desired (this phenomenon can be observed in \textbf{Fig.~\ref{sweep}} as the flat lines from 10 to 20 psf). Therefore, the limiting factor for joint spacing should only be the joint's radius of curvature.

\subsection{Aerodynamic Testing}
\begin{figure}[H]
        \centering
        \includegraphics[width=1.0\linewidth,angle=0]{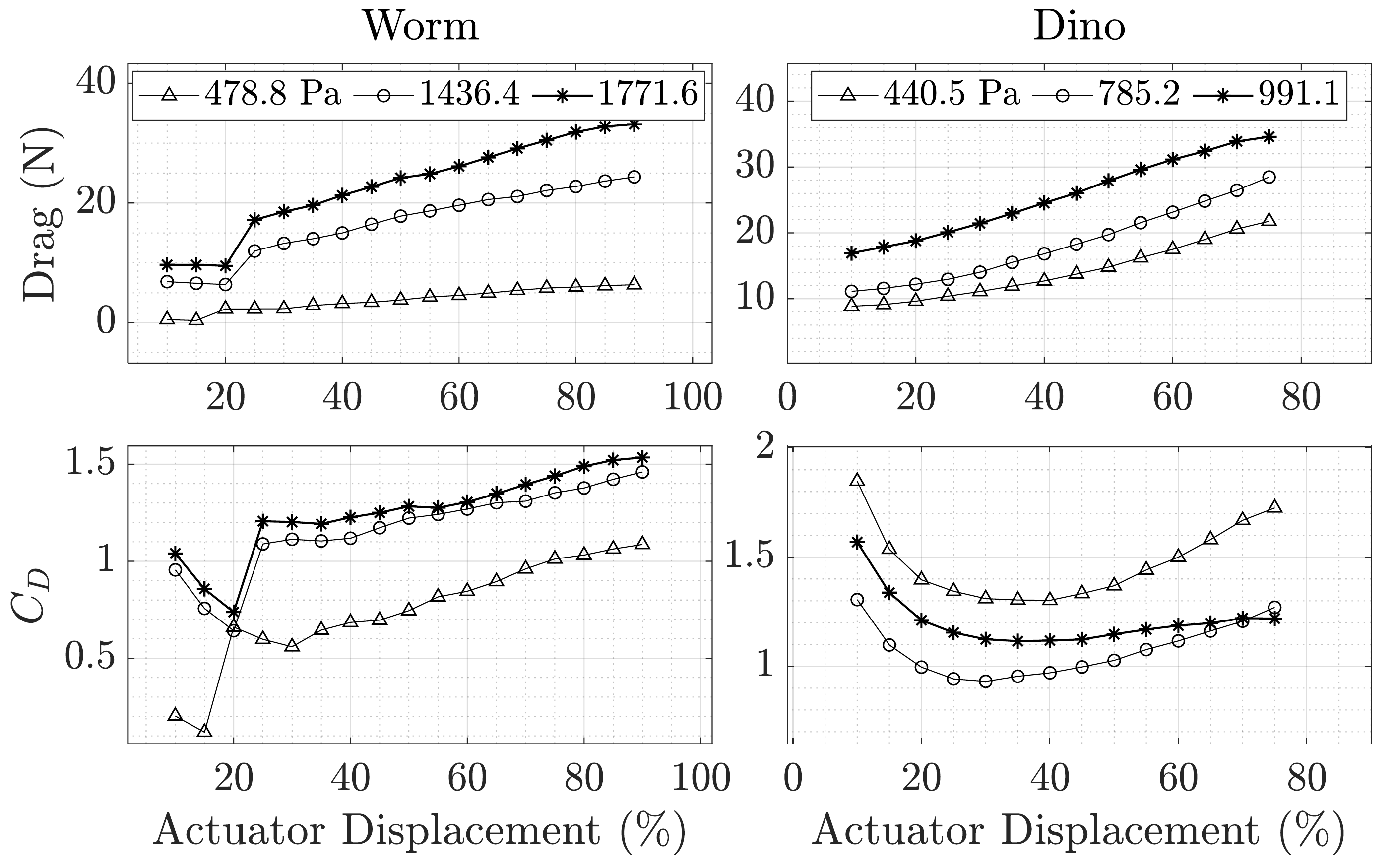}
        \caption{Measured drag and drag coefficients at different dynamic pressures and actuation percentages for the Worm and Dino.}
        \label{sweep}
    \end{figure}

In order to quantify the forces experienced by the test apparatus at various levels of actuation, a number of actuation sweeps were performed in the wind tunnel. \textbf{Fig.~\ref{sweep}} shows drag forces and coefficients determined experimentally in the wind tunnel at dynamic pressures of $478.8$, $1436.4$, and $1771.6 \mbox{ Pa}$ over actuation percentages ranging from 10-90\% for the Worm test assembly, and dynamic pressures of $440.5$, $785.2$, and $991.1$ $\mbox{ Pa}$ over actuation percentages ranging from 10-70\% for the Dino. For calculations of drag coefficients ($C_{D}$), \textbf{Eq.~\ref{Dda_worm}} and \textbf{Eq.~\ref{Dda_dino}} were used. The area of the front plate ($0.0143 m^2$) was taken as the reference area for the Worm, and an area of 0.0434 m$^{2}$ was used for the Dino. Error bars associated with the load cell uncertainty of $\pm$0.1 N were too small to be visible in plots and were therefore neglected.

It is clear to see in \textbf{Fig.~\ref{sweep}} that as actuation approaches 100\% and the front plate of the Worm becomes perpendicular to the free-stream flow, the Worm $C_{D}$ calculated for q = $1436.4$ and $1771.6 \mbox{ Pa}$ approach that of a flat plate ($C_{D} \approx$ 1.2). This result provides confidence in the validity of the results at these dynamic pressures. It is also clear that the Worm $C_{D}$ at low actuation percentages represents a stark departure from the overall trend. The cause of this was due to the flexible joints being too large, impeding the initial actuation as described earlier.

For the Dino, it can be seen that both the drag force and $C_{D}$ trends transition from concave up to concave down at some intermediate actuation percentage. This phenomenon is likely caused by the folding procedure of the Dino. Early on in the folding process, the inner fan transitions from being stowed to fully extended. This accounts for the transition between the initially steeper increase in drag while the fan is being deployed, to a shallower increase once it has already been deployed. This is supported by the surface area trend modeled by \textbf{Eq.~\ref{DinoSurfaceArea}}, where it can be seen that the rate of change of surface area decreases at higher angles of attack, resulting in lower increases in drag generation. It is also interesting to note that the Dino $C_{D}$ for the highest dynamic pressure of $991.1 \mbox{ Pa}$ is actually lower than that for $440.5 \mbox{ Pa}$. There are two possible explanations for this result. First, as the drag coefficient is proportional to (D/q), it is likely the case that drag is increasing at a diminishing rate in proportion to q. This phenomenon is observed in the wind tunnel results for many different rigid bodies. Second, it is also likely that strong wind perturbations were generated off the bow of the test apparatus that had more deleterious effects at increased wind speeds. 

\subsection{Comparison between experimental results and FSI simulations}
In order to verify and validate the FSI simulations used for the design of the Aerogami, the computational results were compared to the experimental data obtained in the wind tunnel. A simulation was run for Worm at $53.6 \mbox{ m/s}$ ($120 \mbox{ mph}$) and 90\% actuation. This was the highest wind speed tested in the wind tunnel and the CFD results showed that Worm would produce $41.2 \mbox{ N}$ of drag. This is in good agreement with initial analytical calculations in which the front panel of Worm was approximated as an inclined flat airfoil and also in good agreement with the wind tunnel experimental setup which yielded $35 \mbox{ N}$ of drag (16.3\% difference). The Dino run at $26.8 \mbox{ m/s}$ ($60 \mbox{ mph}$) at 70\% actuation produced $26.8 \mbox{ N}$ of drag compared to the experimental $21.8 \mbox{ N}$ (20.6\% difference). The fact that the CFD drag was higher is due to the fact that in the wind tunnel the test stand blocks flow coming under the origami. Therefore, the region under the panels would not have such low pressure, and since the pressure under (or behind) the panels is not as low then there would be less pressure drag. When the Worm was run again at $53.6 \mbox{ m/s}$ but this time with the test structure underneath, the output drag from CFD was only $34.1 \mbox{ N}$ of drag (2.6\% difference). The Dino at $26.8 \mbox{ m/s}$ and supporting structure produced $21.3 \mbox{ N}$ of drag (2.3\% difference). The sources of error compared to the wind tunnel testing are: (i) the stand geometry was not modeled in the FSI framework, (ii) friction in the test stand rails would affect load sensor readings but was neglected, and (iii) the temperature and density in the wind tunnel were not standard. 
The excellent agreement between simulations and experimental results obtained in this study shows the potential of high-fidelity computational modeling for the design and topology optimization of composite origami structures.

\subsection{Control Response Testing}
\begin{figure}[H]
        \centering
        \includegraphics[width=0.9\linewidth,angle=0]{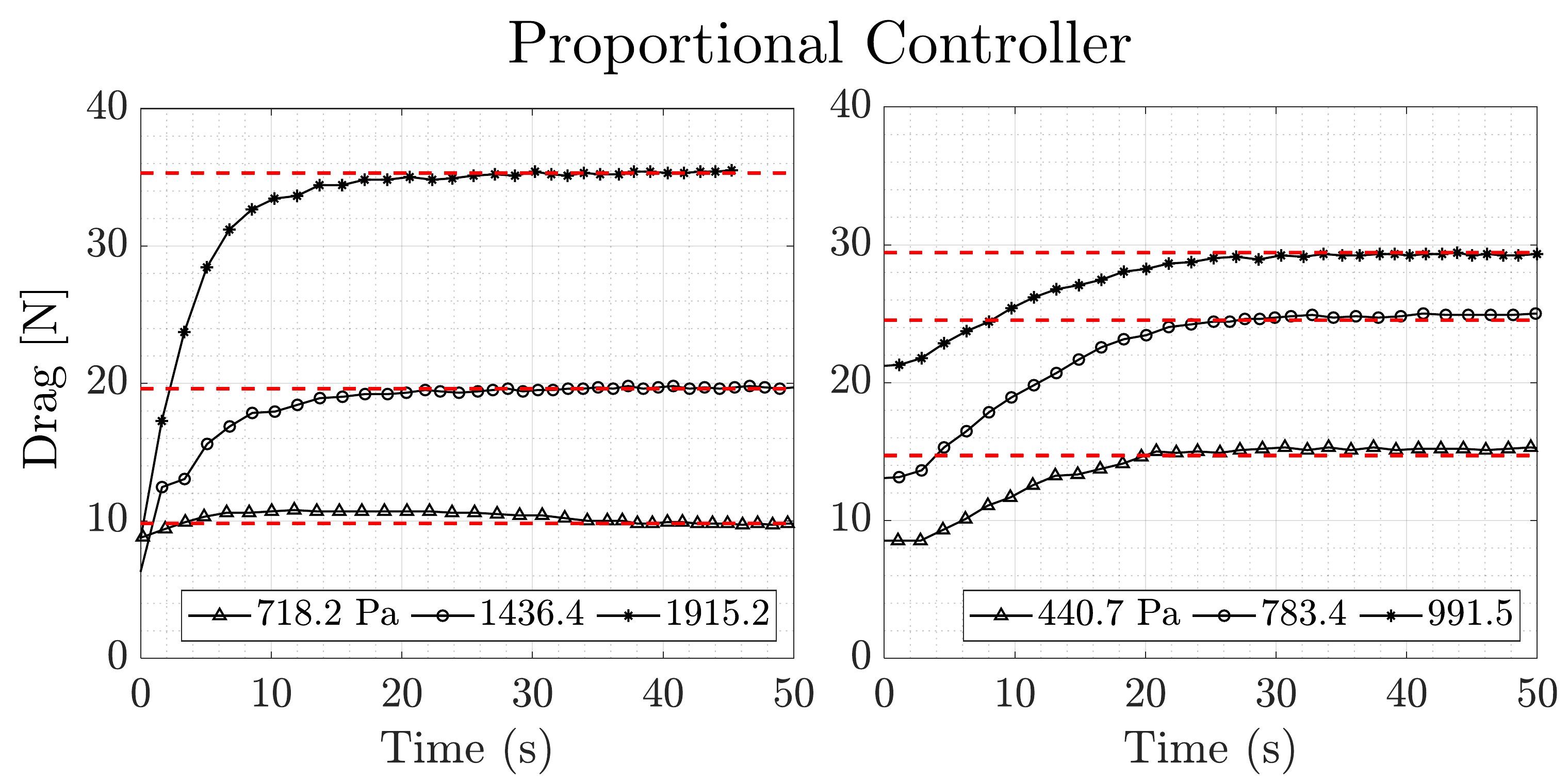}
        \caption{Left: Proportional control law responses for Worm at dynamic pressures of 718.2, 1436.4, and 1915.2 Pa targeting drag values of 9.81, 19.62, and 35.32 N, respectively. Right: Dino at dynamic pressures of 440.7, 783.4, and 991.5 Pa targeting drag values of 14.72, 24.53, and 29.43 N, respectively. Targets are shown by dashed lines.}
        \label{proplaw_responses}
    \end{figure}

\noindent After sweeps had been performed, the proportional control law was tested at dynamic pressures of q = $718.2$, $1436.4$, and $1915.2 \mbox{ Pa}$ targeting drags of $9.81$, $19.6$, and $35.3 \mbox{ N}$ for the Worm. Then the proportional control law was tested on the Dino at dynamic pressures of q = $440.7$, $783.4$, and $991.5 \mbox{ Pa}$ targeting drag forces of $14.7$, $24.5$, and $29.4 \mbox{ N}$, respectively. The results can be seen in \textbf{Fig.~\ref{proplaw_responses}}, where the red dashed lines represent the target drag for each run. It is worth noting that the target drags were varied case-by-case in order to place each targeted drag value towards the mid-range of actuation length at each run to avoid over-extending the actuator.

\noindent The proportional control response behaved as predicted by simulation. As seen in \textbf{Fig.~\ref{proplaw_responses}} the proportional controller produced little to no overshoot, indicating an over-damped control response which converged in all experiments. Near critically damped behavior was originally desired in the control response, but low actuation speeds limited the rise time and required the response to remain in an over-damped regime. It can also be seen that proportional control effectively rejected the signal delay that would cause any oscillations. Finally, it is interesting to note the proportional controller also exhibited effective rejection of large amounts of load measurement fluctuations caused by vibration. Because of these disturbance rejection properties, a proportional controller should be adequate for future tests.

It is worth mentioning that in several applications such as e.g. air brakes, the actuator is usually stiff and the speed of response might not be particularly fast. In this scenario, the control may not be a critical factor. However, the proposed design is intended as a general drag control surface, for which air brakes is only one specific application. Depending on the specific case, the control system can actually play an important role on the performance of the Aerogami. For instance, the use of an Aerogami design is currently under investigation for the active control of the angle of attack in Vertical Axis Wind Turbine (VAWT) blades. In this case, the configuration of the Aerogami must be adapted continuously and the speed of response of the control is key.

\subsection{Future Developments}
The Aerogami concept proposed in this study could represent the basis for the development of novel control surface designs for aeronautical or terrestrial vehicle applications. For instance, Fujiwara \textit{et al.} \cite{fujiwara2017aeroestructural, ting2018optimization} provide optimized aerostructural designs of morphing wings that were proved to reduce drag and improve overall aerodynamic characteristics. Aerogami could be used to enable a continuously morphing trailing edge throughout the wing while reducing the complexity of the actuation mechanisms and the number of components. Early NASA work \cite{freeland1997large, eskenazi2005promising} also showed the possibility of using deployable structures for space exploration. The use of composite Aerogami-type structures \textit{in lieu} of classical metal-alloy systems would provide better performance in terms of weight, strength, durability, and damage tolerance.

\section{Conclusion}
Two origami aerodynamic control surfaces, Worm and Dino, successfully demonstrated the feasibility of constructing deployable drag surfaces with composite origami. Worm successfully functioned as a minimum viable product concept and paved the way for the more complex Dino, which more effectively leveraged the use of origami in the drag surface design. 

Wind tunnel testing confirmed that Worm and Dino can fold correctly past initial actuation percentages with only a single actuator, even at high wind speeds. The experiments revealed excellent controllability of the drag forces over a broad range of deployment stages and aerodynamic pressures for both the designs. 

The experimental results are in very good agreement with Fluid Structure Interaction (FSI) simulations. For Worm, the simulated drag force in a $90\%$ deployment state and a wind speed of $53.6$ m/s ($120$ mph) was found to be $34.1$ N which differs of only $2.6\%$ compared to the experimental value of $35$ N. A similar agreement was found for Dino. For a $70\%$ deployment state and a wind speed of $26.8$ m/s ($60$ mph), the simulations yielded a drag force of $21.3$ N compared to an experimental value of $21.8$ N.

The foregoing results, their repeatability, and the very good agreement between experiments and simulations provide confidence in the feasibility of applying origami composites as aerodynamic control surfaces in real-world applications and utilizing high-fidelity models for the optimization of their topology.

\section*{Acknowledgments}
\noindent The authors would like to express their deep gratitude to the William E. Boeing Department of Aeronautics and Astronautics at the University of Washington (UW-AA) for providing its facilities and resources for the wind tunnel testing. We would also like to thank Prof. James C. Hermanson at the UW-AA for his advice and the insightful discussions on the project.

\newpage
\linespread{1}\selectfont
\bibliographystyle{elsarticle-num}
\bibliography{bib.bib}

\newpage


    \begin{figure}[H]
        \centering
        \includegraphics[width=1\linewidth,angle=0]{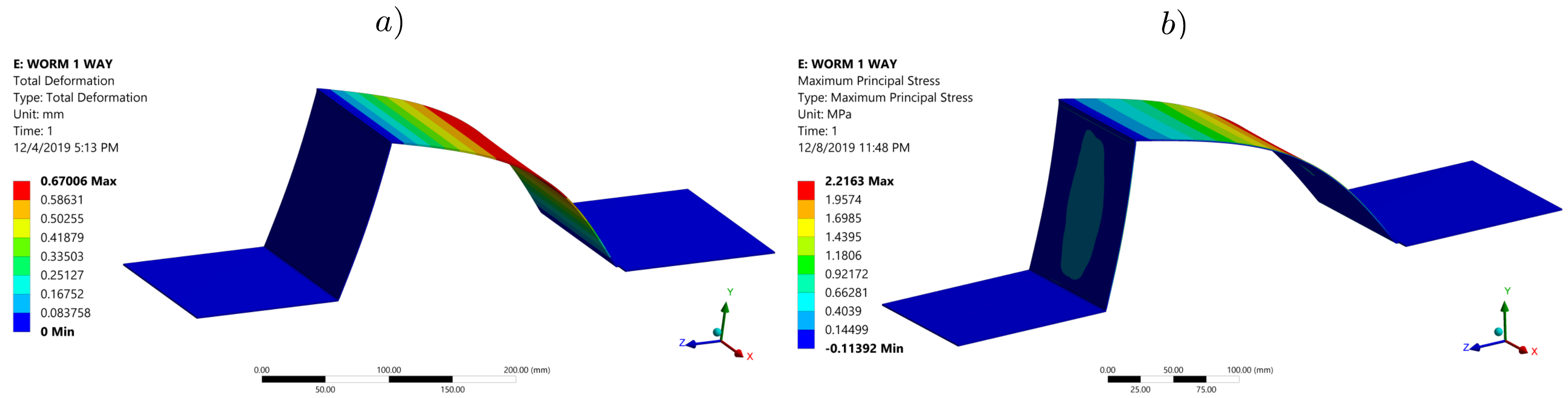}
        \caption{Worm Aerogami Finite Element Analysis results using Ansys Mechanics. a) Total Deformation, b) Maximum Principal Stress}
        \label{fig:FEA_results}
    \end{figure}

    \begin{figure}[H]
        \centering
        \includegraphics[width=1\linewidth,angle=0]{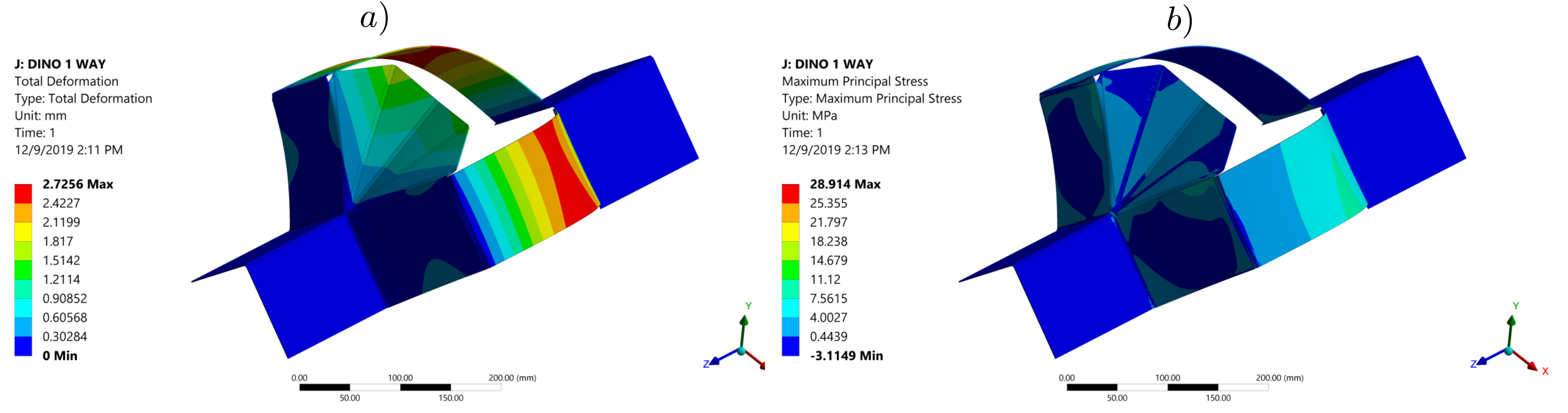}
        \caption{Dino Aerogami Finite Element Analysis results using Ansys Mechanics. a) Total Deformation, b) Maximum Principal Stress}
        \label{fig:FEA_results2}
    \end{figure}

\end{document}